\documentclass{pasj00}

\begin{document}
\SetRunningHead{Saitoh et al.}{Star formation criteria and galactic structures}
\Received{2007/12/04}
\Accepted{2008/02/18}

\title{Toward first-principle simulations of galaxy formation:\\
I. How should we choose star formation criteria in high-resolution simulations of disk galaxies?}

\author{Takayuki \textsc{R.Saitoh}$^1$,
        Hiroshi \textsc{Daisaka}$^2$,
        Eiichiro \textsc{Kokubo}$^{1,3}$,
        Junichiro \textsc{Makino}$^{1,3}$,
        Takashi \textsc{Okamoto}$^4$,\\
        Kohji \textsc{Tomisaka}$^{1,3}$,
        Keiichi \textsc{Wada}$^{1,3}$,
        Naoki \textsc{Yoshida}$^5$ (Project Milkyway)
}
\affil{$^1$ Center for Computational Astrophysics,
        National Astronomical Observatory of Japan, 2--21--1 Osawa,
        Mitaka-shi, Tokyo 181--8588}

\affil{$^2$ Graduate School of Commerce and Management, 
        Hitotsubashi University, Naka 2-1 
        Kunitachi-shi, Tokyo 186--8601}

\affil{$^3$ Division of Theoretical Astronomy,
        National Astronomical Observatory of Japan, 2--21--1 Osawa,
        Mitaka-shi, Tokyo 181--8588; 
        and School of Physical Sciences, Graduate University of Advanced Study (SOKENDAI)}

\affil{$^4$ Institute for Computational Cosmology, Department of Physics, 
        Durham University, South Road, Durham, DH1 3LE, UK}

\affil{$^5$ Department of Physics, Nagoya University, 
        Furocho, Chikusa, Nagoya 464-8602}

\email{saitoh.takayuki@nao.ac.jp,saitoh.takayuki@cfca.jp}


\maketitle

\begin{abstract}

We performed three-dimensional $N$-body/SPH simulations 
to study how mass resolution and other model parameters such as the 
star formation efficiency parameter, $C_*$ and the threshold density
affect structures of the galactic gaseous/stellar disk in a static galactic potential. 
We employ $10^6 - 10^7$ particles to resolve a cold ($T < 100~{\rm K}$) and 
dense ($n_{\rm H} > 100~{\rm cm^{-3}}$) phase as well as diffuse, hot phases.
We found that structures of the interstellar medium (ISM) and the distribution of young 
stars are sensitive to the assumed threshold density for star formation, $n_{\rm th}$.  
High-$n_{\rm th}$ models with $n_{\rm th} = 100~{\rm cm^{-3}}$ yield clumpy 
multi-phase features in the ISM. 
Young stars are distributed in a thin disk of which  
half-mass scale height is $10 - 30~{\rm pc}$.
In low-$n_{\rm th}$ models with $n_{\rm th} = 0.1~{\rm cm^{-3}}$, which is
usually employed in cosmological galaxy formation simulations, 
the stellar disk is found to be several times thicker, 
and the gas disk appears smoother than the high-$n_{\rm th}$ models.
A high-resolution simulation with high-$n_{\rm th}$ 
is necessary to reproduce the complex structure of the
gas disk.
The global properties of the model galaxies in low-$n_{\rm th}$ models,
such as star formation histories, are similar
to those in the high-$n_{\rm th}$ models when we tune the value of $C_*$ so that
they reproduce the observed relation between surface gas density and surface
star formation rate density.
We however emphasize that high-$n_{\rm th}$ models automatically reproduce the
relation, regardless of the values of $C_*$.  In high-$n_{\rm th}$ models, the
difference in star formation histories is within a factor of two for two
runs with values of $C_*$ which differ by a factor of 15.
The ISM structure, phase distribution, and distributions of young star
forming region are also quite similar between these two.
From the analysis of the mass flux on phase diagram,
we found that the timescale of the flow from the reservoir ($n_{\rm H} \sim 1~{\rm cm^{-3}}$) 
to the star forming regions ($n_{\rm H} \gtrsim 100~{\rm cm^{-3}}$) is 
about five times as long as the local dynamical time and 
this evolution timescale is independent of the value of $C_*$.
The use of a high-$n_{\rm th}$ criterion for star formation in high-resolution
simulations makes numerical models fairy insensitive to the modelling of
star formation.

\end{abstract}

\section{Introduction}
A number of physical processes affect 
the formation and evolution of galaxies. 
Star formation is among the most important processes,
not only because it largely determines the bulk properties of a galaxy,
but also because the history of star-formation essentially
reflects the formation history of a galaxy.

Numerical simulation is a powerful tool to study galaxy formation.
To compare ``simulated'' galaxies with observed ones, 
it is necessary to follow the dynamics of baryonic matter as well as 
the assembly of dark matter halos. 
Simulations of galaxy formation are, however, often hampered by 
the fact that relevant physics are still poorly understood.
Numerical resolution is another limiting factor.
In particular, appropriate physical models of star formation should be
used in high resolution simulations.
For example, recent simulations of galaxy formation (e.g., \cite{Governato+2007}),
have a spatial resolution of several hundreds pc, with the corresponding mass 
resolution of $\sim 10^{5}~\Mo$.
In such simulations, 
simple models such as an isothermal interstellar medium (ISM)
are applied to galactic gas disks.
In addition, individual giant molecular clouds (hereafter GMCs) 
in galaxies are not resolved in current simulations,
although GMCs are regarded as the site of star formation.
Thus one often needs to use phenomenological models,
to describe the star formation processes, which is called subgrid physics.

There are many prescriptions (``subgrid models'') of star formation
used in simulations of galaxy formation with coarse resolutions. 
A commonly used technique is to convert high-density 
gas elements to collisionless ``star'' particles 
(e.g., \cite{Katz1992,NavarroWhite1993,
SteinmetzMuller1994,MihosHernquist1994,Katz+1996,
Yepes+1997,ThackerCouchman2001,Abadi+2003,KawataGibson2003,
Sommer-Larsen+2003,SpringelHernquist2003,Robertson+2004,
SaitohWada2004,Okamoto+2005,Stinson+2006,Governato+2007,Okamoto+2007}).
Typical criteria to spawn star particles are as follows 
(e.g., \cite{NavarroWhite1993,Katz+1996,Stinson+2006}):
(1)~the physical density is greater than $0.1~{\rm cm^{-3}}$,
(2)~the temperature is lower than $\simeq 10000~{\rm K}$,
and (3)~the velocity field is converging.
If these three conditions are satisfied,
`stars' are then formed at a rate following the local Schmidt law.
Namely, the local star formation rate (SFR), $d \rho_*/dt$, is 
assumed to be proportional to the local gas density, $\rho_{\rm gas}$, 
and inversely proportional to the local dynamical time, $t_{\rm dyn} \sim 1/\sqrt{G \rho_{\rm gas}}$:
\begin{equation}
\frac{d \rho_*}{dt} = C_* \frac{\rho_{\rm gas}}{t_{\rm dyn}}, \label{eq:sf}
\end{equation}
where $C_{*}$ is the dimensionless star formation efficiency parameter. 
The value of this parameter is usually calibrated by the global star formation properties, 
the Schmidt-Kennicutt relation \citep{Kennicutt1998,MartinKennicutt2001}.
Choosing $C_{*} \sim 0.01$  reproduces 
the Schmidt-Kennicutt relation in the local universe (e.g., \cite{NavarroSteinmetz2000}).
However, the threshold density is too low which 
does not correspond to typical densities of 
the neutral hydrogen (HI) and molecular hydrogen (H$_2$) gas in real galaxies.

There are several models that assumes higher density regions as star forming regions.
For instance, \citet{Kravtsov2003} adopted $n_{\rm H} > 50~{\rm cm^{-3}}$ as the star forming regions 
in a cosmological simulation of galaxy formation.
When he chose a plausible star formation time, 
$t_{\rm sf}$ : $d \rho_{*}/dt = \rho_{\rm gas}/t_{\rm sf}$, 
which is set to constant ($= 4~{\rm Gyr}$) in his simulation,
the Schmidt-Kennicutt relation is also reproduced.
More recently, 
\authorcite{TaskerBryan2006} (\yearcite{TaskerBryan2006,TaskerBryan2007})
performed adaptive mesh refinement simulations of the ISM 
in a static halo potential.
Their simulations resolve individual star forming regions (the minimum cell sizes are $25-50~{\rm pc}$).
They compared two variants of star formation criteria:
(a)~$n_{\rm H} > 10^{3}~{\rm cm^{-3}}$, $T < 10^3~{\rm K}$, and $C_{*} = 0.5$,
and (b)~$n_{\rm H} > 0.02~{\rm cm^{-3}}$, $T < 10^4~{\rm K}$, and $C_{*} = 0.05$.
Both models also employ converging flows as one of the star formation criteria and
the star formation law described by equation (\ref{eq:sf}).
Interestingly, it is found that both the models reproduce the Schmidt-Kennicutt relation.
The star formation histories are also found to be similar.
Therefore, their results appear to imply that the global star formation properties 
are not sensitive to the details of star formation prescriptions.

In this paper, we examine how numerical prescriptions of 
star formation affect structure of the ISM, and the global star formation history (SFH) of a galactic disk.
In particular, we focus on the threshold density in star formation criteria ($n_{\rm th}$)
and the star formation efficiency ($C_{*}$).
We adopt two values of density threshold: $0.1~{\rm cm^{-3}}$ (low-$n_{\rm th}$ model) 
and $100~{\rm cm^{-3}}$ (high-$n_{\rm th}$ model).
We also test the effect of star formation efficiency parameter $C_*$ in high-$n_{\rm th}$ models.
We perform high-resolution SPH simulations (number of SPH particles are $10^6-10^7$) 
in a galactic potential
and we compare structure of the ISM and stellar disks.
We show that, while both models can exhibit similar SFHs and 
the relation of surface gas density to surface SFR, 
only high-$n_{\rm th}$ models have the complex, inhomogeneous, and multiphase ISM.
The ISM has a log-normal like probability density distribution (PDF),
while a model that does not include either star formation and supernova (SN) feedback 
has a power-law like PDF; Star formation and SN feedback distort PDF.
Interestingly, structure of the ISM, stellar disks, and SFRs (SFHs) are 
not simply proportional to $C_*$ in high-$n_{\rm th}$ models.
This is because the mass supply timescale from from the reservoir ($n_{\rm H} \sim 1~{\rm cm^{-3}}$) 
to the star forming regions ($n_{\rm H} \gtrsim 100~{\rm cm^{^3}}$) is $\sim 5~t_{\rm dyn}(n_{\rm H})$ 
and this timescale is not affected by the adopted value of $C_*$.

The plan of this paper is as follows.
We mention the relation between numerical resolution and expressible phase of the ISM in Section 2.
In Section 3, we describe properties of a model galaxy and our numerical methods.
Our results are presented in Section 4.
Summary and discussions are given in Section 5.

\section{Estimate of required resolutions to appropriately model the star formation} \label{sec:est}

We consider regions with the density greater than $100~{\rm cm^{-3}}$ as ``star forming regions''.
From the Jeans condition,
we can estimate the resolution necessary to express the gravitational 
collapse for a fluid with a given density and temperature.
In smoothed particle hydrodynamics (SPH) simulations, 
this condition is expressed as $M_{\rm Jeans} \gtrsim N_{\rm nb} \times m_{\rm SPH}$, 
where $M_{\rm Jeans}$ is the Jeans mass, $N_{\rm nb}$ is a typical number of neighbor particles,
and $m_{\rm SPH}$ is the mass of an SPH particle \citep{BateBurkert1997,Bate+2003,Hubber+2006}.
The Jeans condition then determines resolved regions in phase ($\rho-T$) plane.

\begin{figure}
\begin{center}
\FigureFile(80mm,80mm){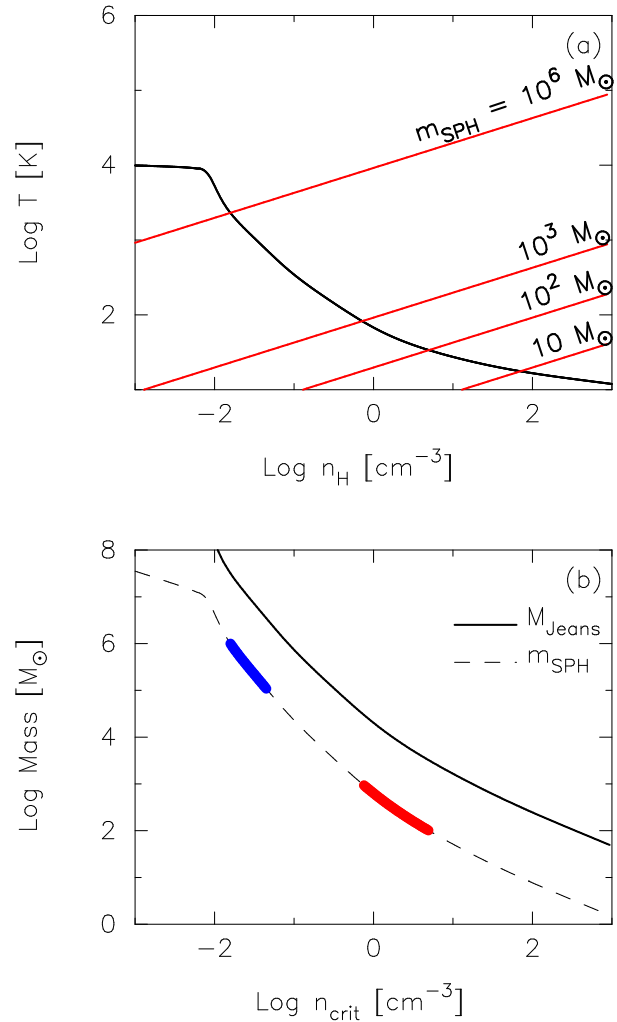}
\end{center}
\caption{Required mass resolutions to resolve cold and dense gas phase (a)
and the critical density as a function of Jeans mass 
in fixed cooling and heating rates using in our simulations (b).
The red lines in panel (a) indicate the Jeans limits (the definition is shown in the text)
with $m_{\rm SPH} = 10^6, 10^{3}, 10^{2}$, and $10~\Mo$ (from top to bottom),
where $m_{\rm SPH}$ is the mass of an SPH particle and we assume that the Jeans mass is $32 \times m_{\rm SPH}$.
The black curve represents the equilibrium temperature of the ISM for
the adopted cooling and heating with the solar abundance, $T_{\rm eq}$. 
For details of the cooling and heating rates, see in \S \ref{sec:darkmatterhalo}.
Panel (b) shows the critical density, $n_{\rm crit}$, 
where $n_{\rm crit}$ is defined as values of the density $n_{\rm crit}$ 
at intersection between equilibrium $\rho-T$ relation 
in panel (a) and constant $M_{\rm Jeans}$ line,
as a function of mass.
Thick solid curve shows $n_{\rm crit}-M_{\rm Jeans}$ relation and 
thin dashed curve represents $n_{\rm crit}-m_{\rm SPH}$ relation.
We assume the mass of each SPH particle is the same.
The red region overplotted on the dashed curve represents 
$n_{\rm crit}$ of simulations with $m_{\rm SPH} = 10^{2-3}~\Mo$ (this study),
while the blue regions overplotted on the solid curve represents 
mass ranges of several simulations of galaxy formation 
($m_{\rm SPH} \sim 10^6~\Mo$ in \cite{Abadi+2003}; $m_{\rm SPH} \sim 10^5~\Mo$ in \cite{Governato+2007}).
}\label{fig:Jeans}
\end{figure}

Figure \ref{fig:Jeans}a shows the limits for several different particle masses.
We can accurately treat the gravitational fragmentation of fluid in the upper region of each line in the phase diagram .
The line is given in equation (3) in \citet{Saitoh+2006}.
The curve indicates the thermal balance between the cooling 
and heating that are both adopted in this paper (see \S \ref{sec:model}).
We assume $N_{\rm nb} = 32$.
It is clear that a very higher resolution is required 
in order to resolve the gravitational collapse to dense clouds, e.g., GMCs.

Figure \ref{fig:Jeans}b illustrates 
$M_{\rm Jeans}$ and $m_{\rm SPH}$ for a given critical density, $n_{\rm crit}$,
where $n_{\rm crit}$ is defined as values of the density $n_{\rm crit}$ 
at intersection between equilibrium $\rho-T$ relation 
in figure \ref{fig:Jeans}a and constant $M_{\rm Jeans}$ line.
The solid curve indicates $M_{\rm Jeans}$ as a function of $n_{\rm crit}$,
while the dashed curve represents $m_{\rm SPH}$ as a function of $n_{\rm crit}$.
The red region overplotted on the dashed curve represents $n_{\rm crit}$ of simulations in this paper,
where the mass range of SPH particles in simulations is 
on the order of $10^{2-3}~\Mo$ (see Table \ref{tab:runs}).
For comparison, we choose several simulations of galaxy formation 
($m_{\rm SPH} \sim 10^6~\Mo$ in \cite{Abadi+2003}; 
$m_{\rm SPH} \sim 10^5~\Mo$ in \cite{Governato+2007})
and plot the corresponding range of the critical density as the blue region.

\section{Methods and models} \label{sec:model}

We investigate 3-D evolution of a gas disk in a static disk-halo potential.
We assume a Navarro-Frenk-White (NFW) density profile \citep{Navarro+1997} for a dark matter halo,
and a Miyamoto-Nagai model \citep{MiyamotoNagai1975} for a stellar disk.
For the halo model, we adopt a cosmological model of 
a standard $\Lambda$ CDM universe \citep{Spergel+2003}.
The cosmological parameters are 
$\Omega_{\rm M} = 0.3$, $\Omega_{\rm \Lambda} = 0.7$, 
and $H_{0} = 70~{\rm km~s^{-1}~Mpc^{-1}}$.
We use these parameters to model the dark matter halo.
By adopting the static potentials for a dark matter halo and a stellar disk, 
we can prevent global disk instabilities and artificial disk heating 
due to the scattering of gas and star particles by massive dark matter particles,
which are inevitable if we use a live halo with low mass resolution.
We will study the effect of live halo and 
also the effect of the mass resolution of halo particles in forthcoming papers.

Gravitational forces are computed by a parallel tree code ASURA 
(\authorcite{ASURA}, in preparation) 
that utilizes the special purpose hardware GRAPE.
The parallel implementation with GRAPE is based on that of \citet{Makino2004}.
We use an opening angle $\theta = 0.5$ for a cell opening criterion.
We only use a monopole moment.
Hydrodynamics is followed by the standard SPH method 
(e.g., \cite{Lucy1977,GingoldMonaghan1977,Monaghan1985,Monaghan1992}). 
The kernel size of each SPH particle is determined by imposing the number of 
neighbors to be $32 \pm 2$.
We use a cooling function for a gas with the solar metallicity 
for a temperature from $10$ K to $10^8$ K \citep{SpaansNorman1997}.
An uniform heating from the far-ultraviolet (FUV) radiation observed 
in the solar neighborhood \citep{Wolfire1995} is included.
We do not include heating from an ultra-violet (UV) background radiation.
This is because we consider the structure of the ISM 
in the current environment of the Milkyway galaxy in this paper.
When we consider the detailed structure of the ISM in the galaxy formation process,
a careful treatment of the UV background and the local FUV radiations is required
since the fluxes are quite stronger than those in the local universe
in high redshift universe.
This is beyond the scope of this paper.

\subsection{Halo+disk galaxy model} \label{sec:darkmatterhalo}

We assume that the dark matter density profile is described by a NFW profile:
\begin{eqnarray}
\rho_{\rm halo}(x) &=& \frac{\rho_{\rm c}}{x (1+x)^2}, x = r/r_{\rm s}, \\ \label{eq:nfw_profile}
c_{\rm NFW} &=& r_{\rm vir}/r_{\rm s}, \\
M_{\rm vir} &=& \frac{4 \pi}{3} \rho_{\rm cr} \Omega_{\rm M} \Delta_{\rm vir} r_{\rm vir}^3,
\end{eqnarray}
where $\rho_{\rm c}$ is the characteristic density of the profile,
$r$ is a distance from the center of the halo,
$r_{\rm s}$ is a scale radius for the profile, 
$\rho_{\rm cr}$ is the critical density of the universe,
and $\Delta_{\rm vir}$ is the virial overdensity 
(we employ $\Delta_{\rm vir} \equiv 340$).
The halo mass is set to be $M_{\rm vir} = 10^{12} \Mo$ and
the concentration parameter is set to be $c_{\rm NFW} = 12$ \citep{Klypin+2002}.
Then the profile has $r_{\rm vir} = 258~{\rm kpc}$, 
$r_{\rm s} = 21.5~{\rm kpc}$, 
and $\rho_{\rm c} = 4.87 \times 10^{6}~{\rm \Mo~kpc^{-3}}$.

The stellar disk is assumed to follow the Miyamoto-Nagai model:
\begin{eqnarray}
\rho_{*}(R,z) &=& \Bigl ( \frac{M_{*} z_{*}^2}{4 \pi} \Bigl ) \times \\ \nonumber
&&\frac{R_{*} R^2 + \bigl ( R_{*}+3\sqrt{z^2+{z_{*}}^2} \bigl ) \bigl ( R_{*}+\sqrt{z^2+{z_{*}}^2} \bigl )^2}
     {\bigl [ R_{*}^2+ \bigl (R_{*}+\sqrt{z^2+{z_{*}}^2} \bigl )^2 \bigl ]^{5/2} \bigl ( z^2 + {z_{*}}^2 \bigl )^{3/2}},
\end{eqnarray}
with mass $M_{*}$, radial scale length $R_{*}$ and vertical scale length $z_{*}$, respectively.
$R$ and $z$ are the cylindrical galactcentric radius and the height, respectively.
Numerical values of model parameters are given in Table \ref{tab:galaxymodel}.

Our model of the gaseous disk is similar to that of \citet{Stinson+2006}.
Initially, the disk has a simple exponential surface density profile. 
The radial scale length of the gas disk, $R_{\rm gas}$, is twice as large as the stellar disk, 
motivated by the observation of \citet{BroeilsRhee1997}.
We truncate the gas disk at $R = 1.5~R_{\rm gas}$.
Note that the truncation radius, $R_{\rm trunc} = 10.5~{\rm kpc}$, is 
close to the edge of the molecular disk in the Milky Way galaxy ($\sim 11~{\rm kpc}$; \cite{NakanishiSofue2006}).
The initial vertical distribution follows a Gaussian distribution 
with a scale height of the distribution equals to $z_{*}$ (see Table \ref{tab:galaxymodel}).
The total gas mass is $3.5\times10^9~\Mo$. 
Since we initially place $10^{6-7}$ particles in each run,
each SPH particle has a mass of $350 - 3500~\Mo$ 
(Results of convergence tests are shown in \S \ref{sec:resolution}).
The gas disk initially rotates with a circular velocity of the model galaxy.
The gravitational softening length is set to be $10~{\rm pc}$
for the gas particles.
The initial gas metallicity is set to be the solar value.

We evolve the disk without radiative cooling for the first 50 Myr,
in order to have a relaxed particle distribution.
We have checked that our choice of the time duration of 
this initial relaxation phase does not affect the main results.

\begin{table*}[thb]
\begin{center}
\caption{Parameters of the model galaxy}\label{tab:galaxymodel}
\begin{tabular}{cccccccccccc}
\hline\hline
\multicolumn{2}{c}{DM halo} & & \multicolumn{3}{c}{Stellar disk} & & \multicolumn{5}{c}{Gas disk} \\
\cline{1-2} \cline{4-6}  \cline{8-12} 
${M_{\rm vir}}^{\rm a}$ & ${c_{\rm NFW}}^{\rm b}$ & & ${M_{\rm *}}^{\rm c}$ & ${R_{\rm *}}^{\rm d}$ & ${z_{\rm *}}^{\rm e}$ & &
${M_{\rm gas}}^{\rm f}$ & ${R_{\rm gas}}^{\rm g}$ & ${R_{\rm trunc}}^{\rm h}$ &${z_{\rm gas}}^{\rm i}$ & ${T_{\rm init}}^{\rm j}$\\
\hline
$10^{12} \Mo$ & 12 & & $4.0 \times 10^{10} \Mo$ & $3.5~{\rm kpc}$ & $400~{\rm pc}$ & &
$3.5 \times 10^9 \Mo$  & $7~{\rm kpc}$ & $10.5~{\rm kpc}$ & $400~{\rm pc}$& $10^4~{\rm K}$\\
\hline
\end{tabular} \\
$^{\rm a}$Virial mass of halo ($\Mo$).
$^{\rm b}$Concentration parameter.
$^{\rm c}$Mass of stellar disk ($\Mo$).
$^{\rm d}$Scale length of stellar disk (kpc).
$^{\rm e}$Scale height of stellar disk (kpc).
$^{\rm f}$Mass of gas disk ($\Mo$).
$^{\rm g}$Scale length of gas disk (kpc).
$^{\rm h}$Truncation radius of gas disk (kpc).
$^{\rm i}$Initial scale height of gas disk (pc).
$^{\rm j}$Initial temperature of gas disk (K).
\end{center}
\end{table*}

\subsection{Star formation and supernova feedback models}
We adopt a commonly used condition for star formation:
(1) $n_{\rm H} > n_{\rm th}$,
(2) $T < T_{\rm th}$,
and (3) $\nabla \cdot v < 0$,
for a star formation site.
We parameterize the star formation model by two parameters: $n_{\rm th}$ and $T_{\rm th}$.
Here we consider two simple models for star formation.
One is $n_{\rm th} = 100~{\rm cm^{-3}}$ and $T_{\rm th} = 5000~{\rm K}$.
This density corresponds to mean densities of GMCs,
while this temperature is much higher than the typical temperature of GMCs ($T < 100~{\rm K}$).
Nonetheless, we find that more than ninety-percent of stars in mass 
are formed from the gas below $T = 100~{\rm K}$.
The disk structures are thus insensitive to the choice of $T_{\rm th}$.
We call this model the ``high-$n_{\rm th}$ model''.
The other is $n_{\rm th} = 0.1~{\rm cm^{-3}}$ and $T_{\rm th} = 15000~{\rm K}$.
We dub this model the ``low-$n_{\rm th}$ model''.
The low-$n_{\rm th}$ model is similar to what is used in 
previous simulations of galaxy formation 
(e.g., \cite{NavarroWhite1993,Katz+1996,ThackerCouchman2001,
Okamoto+2005,Governato+2007,Okamoto+2007}).

When an $i$-th gas particle is eligible to form stars, 
we compute the probability $p_{{\rm SF},i}$ of the particle to spawn 
a new star particle with mass ${m_{\rm *,spawn}}$ during a time-step width $dt$ as
\begin{equation}
p_{{\rm SF},i} = \frac{m_{{\rm gas},i}}{m_{\rm *,spawn}}
\Bigl [ 1-\exp \Bigl (-C_{*} \frac{dt}{t_{{\rm dyn},i}} \Bigl )\Bigl ], \label{eq:prop_spawn}
\end{equation}
where $m_{{\rm gas},i}$ is the mass of the gas particle, 
and $t_{{\rm dyn},i} = 1/\sqrt{4 \pi G \rho_{{\rm gas},i}}$, respectively.
If we use $ m_{\rm *,spawn} = m_{\rm gas,i}$, masses of the gas particles around star forming regions become
heavier by receiving mass from evolved stars and we loose mass resolution.
On the other hand, too small value of $m_{\rm *,spawn}$ is
not favored from a dynamical point of view.
We thus fix $m_{\rm *,spawn}$ to one-third of the original gas particle mass
as in \authorcite{Okamoto+2003} (\yearcite{Okamoto+2003}, \yearcite{Okamoto+2005}).
When the mass of a gas particle becomes smaller than $m_{\rm *,spawn}$, 
we convert the gas particle into a collisionless particle.
We consider each stellar particle as a single stellar population (SSP) having its own age and metallicity.
We assume the Salpeter IMF \citep{Salpeter1955} whose lower and upper mass 
limits are $0.1~\Mo$ and $100~\Mo$, respectively.

We implement SN feedback in a probabilistic manner as in \citet{Okamoto+2007}.
We assume that stars more massive than $8~\Mo$ explode as Type II SNe
and each Type II SN outputs $10^{51}~{\rm ergs}$ of thermal energy into the ISM around the SN.
In this paper, we only consider the effect of Type II SNe as feedback from stellar particles,
since the time integration of each run is done only $0.3 - 1~{\rm Gyr}$ and
the lifetime of Type Ia SNe progenitor is $\gtrsim {\rm Gyr}$.
The number of SNe in each SSP is approximated by a single event.
The probability of a SSP $i$ having such event of SN explosion 
during a time interval $dt$ is given by 
\begin{equation}
p_{{\rm SNII},i} = \frac{\int^{t_{{\rm SSP},i}+dt}_{t_{{\rm SSP},i}} r_{\rm SNII}(t') dt'}
{\int^{t_8}_{t_{{\rm SSP},i}} r_{\rm SNII}(t') dt'},
\end{equation}
where $t_{{\rm SSP},i}$ is the age of the SSP, $r_{\rm SNII}$ is the SN II rate for the SSP, 
and $t_8$ is the lifetime of a $8~\Mo$ star.
SN energy is smoothly distributed over the surrounding 32 SPH particles.
We here use the SPH kernel as a weighting function for the energy deposition.
The specific SN rate is $\simeq 0.0072~{\rm SN}/\Mo$ and 
the typical stellar mass of each stellar particle is $\sim 1000~\Mo$
in our simulation with $10^6$ particles.
Thus each SN event in our simulation corresponds 
approximately to an association of $\simeq 7~{\rm SNe}$.

Table \ref{tab:runs} shows the model parameters for our runs.
We perform a total of seven runs, and label them `A', `B', `C', `D', and `E'
with additional two runs with extra suffixes, such as 3 and 10.
Labels indicate the adopted ranges of radiative cooling function and star formation model.
Run A represents a standard model, which is often used 
in cosmological simulations of galaxy formation.
This model employs $10^4~{\rm K}$ as the minimum temperature, $T_{\rm cut}$.
Then this model does not have the cold phase gas ($T < 1000~{\rm K}$).
In contrast, Run B adopts $T_{\rm cut} = 10~{\rm K}$ and has the cold phase gas.
Both Runs C and D adopt a high-density and a low-temperature thresholds, 
whereas these models have different star formation efficiencies.
Runs C and D employ $T_{\rm cut} = 10~{\rm K}$.
The low star formation efficiency in Run C is motivated by 
the slow star formation model of \citet{ZuchermanEvans1974} and \citet{KrumholzTan2007}, 
whereas the high star formation efficiency in Run D is motivated by 
the observations of star clusters ($C_* \sim 0.1-0.3$) reported by \citet{LadaLada2003}.
Such high efficiency is also adopted in recent simulations 
by \authorcite{TaskerBryan2006} (\yearcite{TaskerBryan2006,TaskerBryan2007})
($C_* = 0.5$).
Runs C$_{3}$ and C$_{10}$ differ from Run C in the mass resolution.
Run E excludes the star formation process.
We evolve Runs A, B, C, D, and E for $1~{\rm Gyr}$, 
whereas we terminate Runs C$_{3}$ and C$_{10}$ at $0.3~{\rm Gyr}$ 
because of limitations of computational resources.

\begin{table*}
\begin{center}
\caption{Parameters of runs}\label{tab:runs}
\begin{tabular}{lccccccc}
\hline
\hline
model & $N^{\rm a}$ & ${m_{\rm SPH}}^{\rm b}$ & ${\epsilon}^{\rm c}$ & 
${T_{\rm cut}}^{\rm d}$ & ${n_{\rm th}}^{\rm e}$ & ${T_{\rm th}}^{\rm f}$ & ${C_{*}}^{\rm g}$ \\
\hline
Run A            & $10^6$ & $3500~\Mo$ & $10~{\rm pc}$ & $10000~{\rm K}$ & $0.1~{\rm cm^{-3}}$ & 15000 K & 0.033 \\
Run B            & $10^6$ & $3500~\Mo$ & $10~{\rm pc}$ & $10~{\rm K}$ & $0.1~{\rm cm^{-3}}$ & 15000 K & 0.033 \\
Run C            & $10^6$ & $3500~\Mo$ & $10~{\rm pc}$ & $10~{\rm K}$ & $100~{\rm cm^{-3}}$ & 5000 K & 0.033 \\
Run C$_{\rm 3}$  & $3 \times 10^6$ & $1170~\Mo$ & $10~{\rm pc}$ & $10~{\rm K}$ & $100~{\rm cm^{-3}}$ & 5000 K & 0.033 \\
Run C$_{\rm 10}$ & $10^7$ & $350~\Mo$ & $10~{\rm pc}$ & $10~{\rm K}$ & $100~{\rm cm^{-3}}$ & 5000 K &0.033 \\
Run D            & $10^6$ & $3500~\Mo$ & $10~{\rm pc}$ & $10~{\rm K}$ & $100~{\rm cm^{-3}}$ & 5000 K & 0.5 \\
Run E            & $10^6$ & $3500~\Mo$ & $10~{\rm pc}$ & $10~{\rm K}$ & N/A & N/A & N/A \\
\hline
\end{tabular}\\
$^{\rm a}$The initial number of SPH particles.
$^{\rm b}$Mass of individual SPH particles ($\Mo$).
$^{\rm c}$Gravitational softening length (pc).
$^{\rm d}$Cut off temperature of cooling function (K).
$^{\rm e}$Threshold density of star formation (${\rm cm^{-3}}$).
$^{\rm f}$Threshold temperature of star formation (K).
$^{\rm g}$Star formation efficiency.
\end{center}
\end{table*}

\section{Results}
We present the three-dimensional structure of the ISM and the distribution of young stars
in \S \ref{sec:Features}.
The SFHs and the relation between surface gas density and surface SFR density
($\Sigma_{\rm gas}-\Sigma_{\rm SFR}$) are shown in \S \ref{sec:starformation}.
Phase ($\rho-T$) diagrams and density probability distribution functions (PDFs) are shown 
in \S \ref{sec:phase}.
The results of the convergence tests are reported in \S \ref{sec:resolution}. 

\subsection{Features of disks}
\label{sec:Features}

\subsubsection{Structure of gas disks}
\label{sec:gasdisks}

\begin{figure*}
\begin{center}
\FigureFile(160mm,160mm){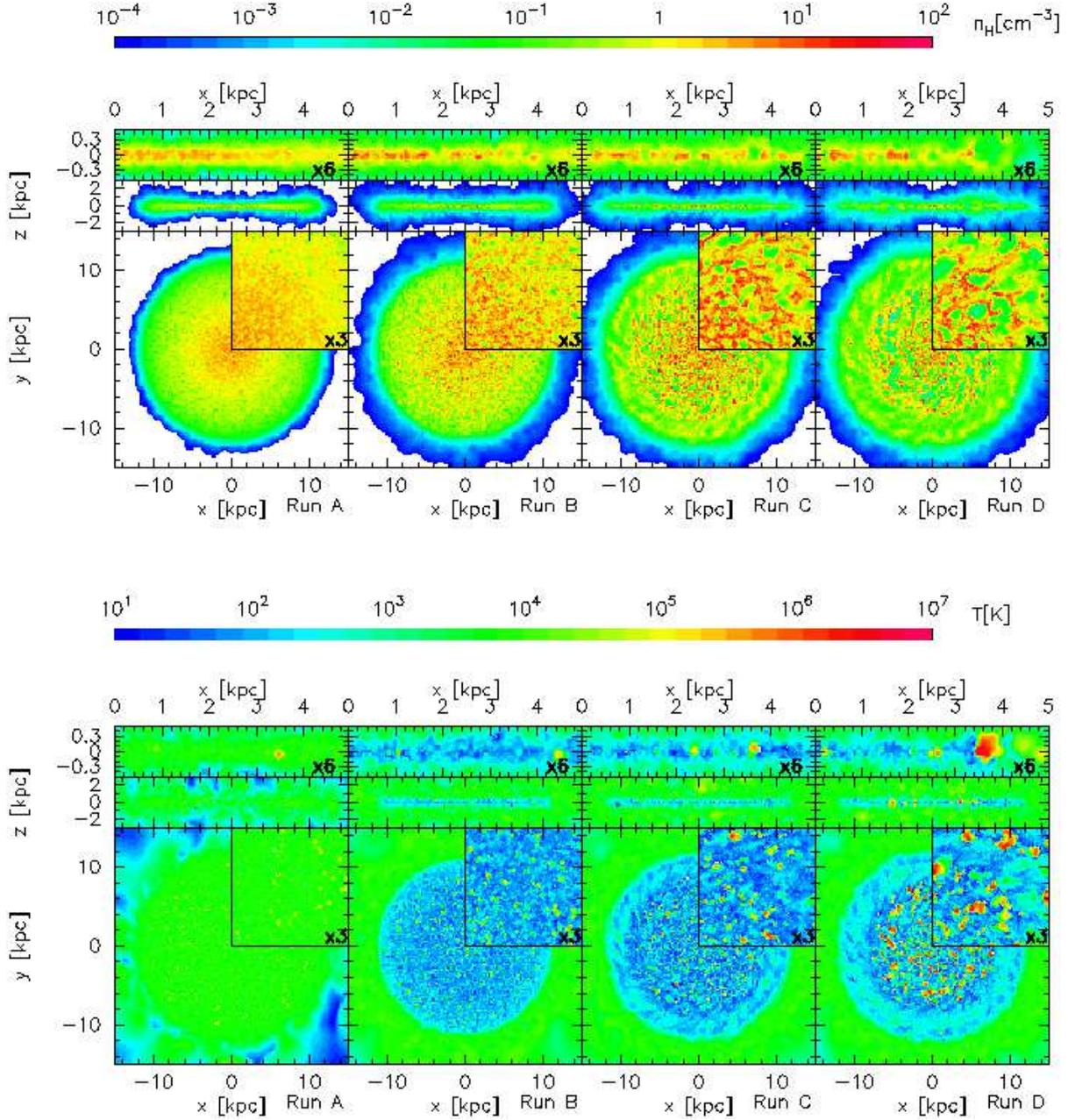}
\end{center}
\caption{
Density (top row) and temperature (bottom row) snapshots 
of Runs A, B, C, and D (from left to right), at $t = 0.3~{\rm Gyr}$.
Top and middle panels show edge-on views (a thin slice of $y = 0$)
while the bottom panel shows a face-on view (a thin slice of $z = 0$).
The middle panels show a region of $-15 < x < 15~{\rm kpc}$ and $-2.5~{\rm kpc} < z < 2.5~{\rm kpc}$.
The top panels show a region of $0 < x < 5~{\rm kpc}$ and $-0.42~{\rm kpc} < z < 0.42~{\rm kpc}$.
The plot range of the bottom panels is $-15 < x < 15~{\rm kpc}$ and $-15 < y < 15~{\rm kpc}$.
The insets in the bottom panels give the close up view of the inner disk for the first quadrant
($0 < x < 5~{\rm kpc}$ and $0 < y < 5~{\rm kpc}$).
}\label{fig:RunsABCD}
\end{figure*}

Figure \ref{fig:RunsABCD} shows density and temperature snapshots 
of Runs A, B, C, and D at $t = 0.3~{\rm Gyr}$.
Different values of $T_{\rm cut}$ lead different density and temperature structures;
the models with $T_{\rm cut} = 10~{\rm K}$ show gas disks with complex and inhomogeneous structures (Runs B, C, and D),
whereas the model with $T_{\rm cut} = 10000~{\rm K}$ show a much smooth gas disk (Run A).
Different star formation criteria yield further different density and temperature structures.
The high-$n_{\rm th}$ models have more complex and inhomogeneous structures 
in the ISM (Runs C and D ) than that in the low-$n_{\rm th}$ model with $T_{\rm cut} = 10~{\rm K}$ (Run B).

Run A has a smooth density structure
because the higher effective pressure ($T_{\rm cut} = 10000~{\rm K}$) stabilizes the gas disk.
In contrast, Run B has cold clumps because of 
the lower temperature threshold for star formation, $T_{\rm cut} = 10~{\rm K}$.
Similar influences of $T_{\rm cut}$ on the ISM structure is also reported in \citet{Saitoh+2006}.
In Runs A and B, the ISM forms stars at local density peaks 
before they develop clumpy structures such as dense filaments,
which are found in high-$n_{\rm th}$ models.

Runs C and D have highly inhomogeneous structures compared with Runs A and B.
In these runs, numerous cold and dense
($T \le 1000~{\rm K}$ and $n_{\rm H} > 10~{\rm cm^{-3}}$)
gas clumps are formed because of the high-$n_{\rm th}$.
We can see many filaments and clumps of dense gases as well as `holes' of diffuse gases.
The dense filaments and clumps have $3-4$ orders of magnitudes larger densities than the holes.
Such structures are rapidly formed after turning on radiative cooling
and they are retained throughout the evolution.
Close comparison of Runs C and D reveals that 
the total mass in dense gas clumps is larger in Run C than in Run D. 
The volume which is filled with dense gas in Run C also appears larger than in Run D.
This is partly because the output time for Run D is just after a star formation peak 
(see figure \ref{fig:SFH}).

From edge-on views of density snapshots in figure \ref{fig:RunsABCD},
we find that the gas disks with $n_{\rm H} > 1~{\rm cm^{-3}}$ have 
a thickness of $\sim 200~{\rm pc}$ for all the models.
In Run A, the disk has a smooth vertical structure.
In Runs B, C, and D, the disks appear rather clumpy, similarly to the results of 
three-dimensional Eulerian simulations for galactic disk 
(e.g., \cite{deAvillez2000,deAvillez2000b,TaskerBryan2006,WadaNorman2007}).
The dense clumps have a low temperature ($T < 1000~{\rm K}$) in these runs.

The temperature distributions are shown in the bottom row of figure \ref{fig:RunsABCD}.
Again, Run A has a smooth and high temperature distribution, 
while Runs B, C, and D have complex structures of cold gas.
It is intriguing that cold components with $T < 1000~{\rm K}$ 
in Runs B, C, and D are not always associated with dense gas layers 
(see close up views of density and temperature maps from edge-on).
Dense gas layers are heated up by hydrodynamic shocks caused by gravity and SNe.
As a result, the temperature at a given density can be different 
from the equilibrium value.
This result is similar to that of previous ISM simulations of galactic disks and circumnuclear disks
(e.g., \cite{WadaNorman1999,WadaNorman2001,WadaTomisaka2005,TaskerBryan2006}).

\begin{figure}
\begin{center}
\FigureFile(80mm,80mm){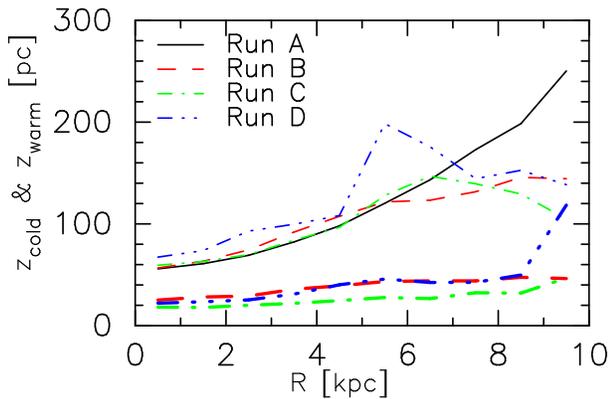}
\caption{Half mass heights of gas disks with
$T < 100~{\rm K}$ and $100~{\rm K} <T < 10000~{\rm K}$ at $t = 0.3~{\rm Gyr}$.
The thick lines represent $z_{\rm cold}$, while the thin curves represent $z_{\rm warm}$.
Solid (black), dashed (red), dot-dash (green), and dash-dot-dot-dot (blue) curves 
indicate Runs A, B, C, and D, respectively.
The scale of cold gas disk ($z_{\rm cold}$) for Run A is not plotted on this figure.
}\label{fig:zrgas}
\end{center}
\end{figure}

We define a characteristic scale height for each phase as the height that
contains half of the mass of the phase at given radius.
We consider two phases, a cold phase ($10~{\rm K} < T < 100~{\rm K}$)
and a warm phase ($100~{\rm K} <T < 10000~{\rm K}$) and 
call their characteristic scale heights, $z_{\rm cold}$ and $z_{\rm warm}$, respectively.
Figure \ref{fig:zrgas} shows $z_{\rm cold}$ (Runs B, C, and D) and $z_{\rm warm}$ (Runs A, B, C, and D)
as a function of $R$ at $t = 0.3~{\rm Gyr}$.
Immediately, we find that typical vertical distribution is clearly separated by gas temperature and 
the distribution is not affected by criteria of star formation in our simulations.
In Runs B, C, and D, $z_{\rm cold}$ is $20-50~{\rm pc}$, whereas $z_{\rm warm}$ is $60-120~{\rm pc}$.
Run D has two local peaks in the profiles: 
one is shown in the curve of $z_{\rm warm}$ at $R \simeq 5~{\rm kpc}$
and another one is shown in that of $z_{\rm cold}$ at $R > 8~{\rm kpc}$.
These two peaks are induced by the localized SNe and are temporary structure.
In Run A, $z_{\rm warm}$ is identical to the others, 
except at the large radii ($R > 7~{\rm kpc}$).
Hence the vertical structures of the ISM are basically determined 
by the ranges of the cooling function.
Both $z_{\rm cold}$ and $z_{\rm warm}$ in all runs gradually increase with increasing $R$.
In Run A, the curve of $z_{\rm warm}$ becomes four times thicker at the edge than that at center.
In Runs B, C, and D, the curves of  $z_{\rm cold}$ and $z_{\rm warm}$
become two times thicker at the edge than those at centers.

\subsubsection{Structure of stellar disks}
\label{sec:stellardisks}

\begin{figure*}
\begin{center}
\FigureFile(160mm,80mm){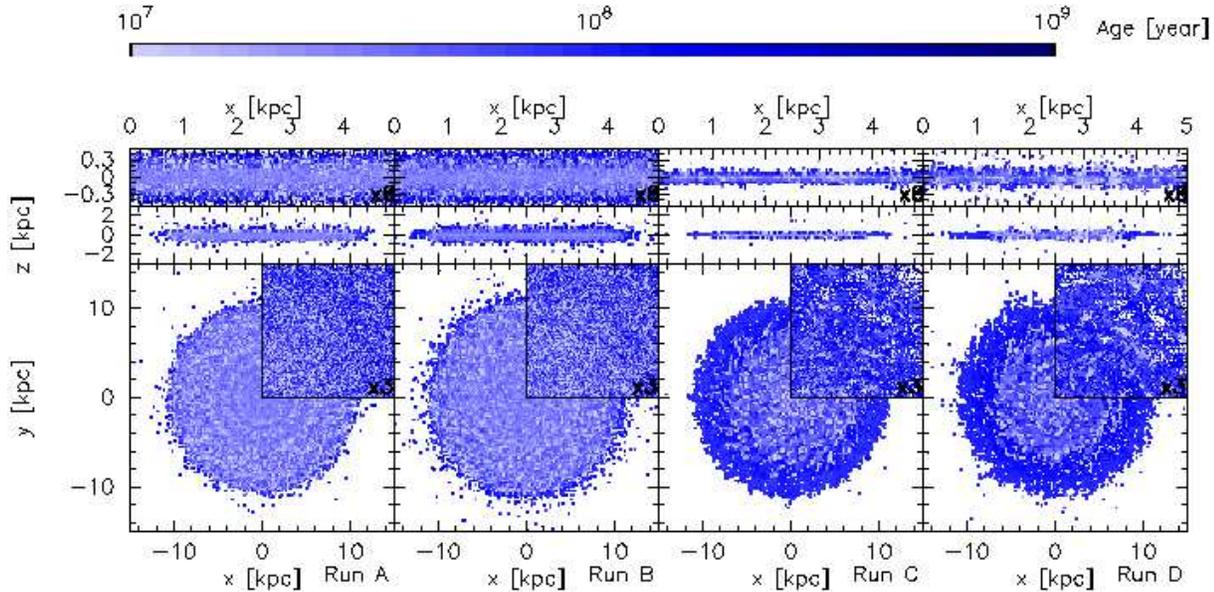}
\end{center}
\caption{Projected star particle distributions in face-on and edge-on views 
of Run A, B, C, and D.
Star particles are assigned on uniform grids.
Each grid size is $234~{\rm pc} = (30~{\rm kpc}/128)$ for for middle and bottom panels,
whereas that is $39~{\rm pc}$ for top panels and the insets in bottom panels.
The color level calculated based on the age of the ten-percent youngest star particles in each grid.
Arrangement of panels is the same as figure \ref{fig:RunsABCD}.
} \label{fig:RunsABCDStars}
\end{figure*}

Figure \ref{fig:RunsABCDStars} shows 
the face-on and edge-on views of the distribution of star particles for the four runs.
We color the figures based on the age of the ten-percent youngest star particles
in each grid as a representative value.
The face-on views clearly show that the distributions of stars 
in Runs C and D have clumpy structures while Runs A and B have smooth structures.
The localization of young stars in high-$n_{\rm th}$ runs 
is easily understood by the complex and inhomogeneous structure of the gas disk.
Close comparison between figures \ref{fig:RunsABCD} and \ref{fig:RunsABCDStars}
reveals that young stars are not always associated with dense gas regions.
This is because SN feedback blows out remaining gas around young stars.
Consequently, no clear spatial correlation is found 
between the distributions of young stars and that of dense gas clumps at any given time.

The edge-on views in figure \ref{fig:RunsABCDStars} suggest that 
the vertical distribution of star particles is strongly affected 
by the threshold density for star formation.
This result indicates that $n_{\rm th}$ is an essential parameter that determines 
the thickness of stellar disk.
We argue that the threshold density needs to be determined by the physical 
value of the star forming regions, such as typical density of molecular clouds.

We further study the vertical structures of the stellar disks.
We define characteristic scale heights as the heights 
that contains 50\%, $z_{*,50}$, and 90\%, $z_{*,90}$,
of the stellar mass at a given radius.
Figure \ref{fig:zrgas} shows $z_{{\rm gas},50}$ and $z_{{\rm gas},90}$ 
as a function of $R$ at $t = 0.3~{\rm Gyr}$.
This indicates clearly that the threshold density for the star formation strongly affects  
the vertical structure of stellar disks.
The low-$n_{\rm th}$ models (Runs A and B) have thick, extended stellar disks
($30-110~{\rm pc}$ for $z_{*,50}$ and  $100-300~{\rm pc}$ for $z_{*,90}$),
whereas the high-$n_{\rm th}$ models (Runs C and D) have very thin stellar disks
($10-30~{\rm pc}$ for $z_{*,50}$ and $30-60~{\rm pc}$ for $z_{*,90}$).
We also find that heights of all stellar disks increase, even weakly, with $R$.
The threshold density has a critical role for stellar disk formation.
In contrast, we note that the value of $C_*$ does not significantly 
affect the vertical structures of stellar disks.

\begin{figure}
\begin{center}
\FigureFile(80mm,80mm){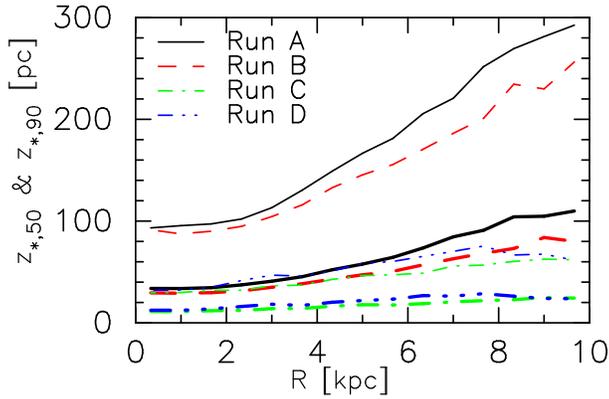}
\caption{
Half mass and ninety-percent mass heights of stellar disks at $t = 0.3~{\rm Gyr}$.
The thick lines represent $z_{*,50}$, while the thin curves represent $z_{*,90}$.
Solid (black), dashed (red), dot-dash (green), and dash-dot-dot-dot (blue) curves 
indicate Runs A, B, C, and D, respectively.
}\label{fig:zrstar}
\end{center}
\end{figure}

\subsubsection{Disk scale heights: comparison of simulations and observations}

In this subsection, we compare the vertical scale heights 
of three components in our models with observations,
namely the vertical distributions of HI and H$_2$ gas disks, and 
young star forming regions (galactic young open clusters).
To simplify the comparison, 
we utilize half-mass heights both simulations and observations.

First, we compare $z_{\rm warm}$ with the half-mass height of the galactic HI gas disk,
since the typical temperature of HI gas is $100~{\rm K} < T < 10000~{\rm K}$ 
\citep{Myers1978,Spitzer1978book}.
\citet{NakanishiSofue2003} reconstructed 
the three dimensional structure of HI gas of the Milky Way galaxy,
by compiling three HI survey data:
the Leiden/Dwingelloo survey \citep{HartmannBurton1997}, Parkes survey \citep{Kerr+1986},
and NRAO survey \citep{BurtonLiszt1983}.
They obtained the vertical scale height, 
which is defined as the full width at half maximum (FWHM) as the vertical scale height, 
as a function of $R$ (see figure 4 in \cite{NakanishiSofue2003}).
To multiple $0.625$ for the FWHM, we obtain the half-mass scale height,
since they assume the vertical distribution of HI gas is proportional to 
the square of a function of hyperbolic secant.
The observationally suggested half-mass scale height ($60-180~{\rm pc}$ at $R = 0 - 10~{\rm kpc}$) is 
almost identical to $z_{\rm warm}$ for the four runs (Runs A, B, C, and D).

Second, we compare $z_{\rm cold}$ and 
the half-mass scale height of the galactic H$_{\rm 2}$ gas disk.
By using the compilation data of $\atom{CO}{}{12}(J = 1-0)$, 
which is provided by \citet{Dame+2001}, 
\citet{NakanishiSofue2006} found that 
the scale heights of the H$_{\rm 2}$ disk (the FWHM of the H$_{\rm 2}$ gas disk)
in the Milky Way galaxy is $48 - 160~{\rm pc}$ at $R = 0 - 11~{\rm kpc}$.
Again, we convert the FWHM to the half mass scale.
The half mass scale of the observed H$_{\rm 2}$ gas disk 
is $\sim 30 - 100~{\rm pc}$ at $R = 0 - 11~{\rm kpc}$.
The half-mass scale heights of simulations, $z_{\rm cold}$, are 
slightly thiner than that in the Milky Way galaxy.
However the difference is at most a factor of two.
Then we consider that $z_{\rm cold}$ is in good agreement 
with that in the Milky Way galaxy.

Finally, we compare the vertical scales of star particles 
with the observed one through mass fractions.
The vertical distribution of galactic young open clusters
is a good tracer of the recent star forming regions in the Milky Way galaxy.
As shown in \citet{JanesPhelps1994}, the vertical distribution 
of galactic young clusters of which
the ages are shorter than that of Hyades ($\sim 800~{\rm Myr}$)
is fitted by an exponential function with a 55-pc scale-height.
The half-mass height of the exponential profile is $\simeq 38~{\rm pc}$ ($0.69 \times 55~{\rm pc}$).
Star particles in high-$n_{\rm th}$ models (Runs C and D) have 
$z_{*,50} \sim 10-30~{\rm pc}$,
whereas stars in the low-$n_{\rm th}$ models (Runs A and B) are more broadly distributed 
in the vertical direction, $z_{*,50} \sim 60-100~{\rm pc}$.
Only the high-$n_{\rm th}$ models are consistent with the observation.
It should be noted that even in Run B which includes the cooling 
under $10^4~{\rm K}$, the scale height is too large.
We can conclude that $n_{\rm th}$ is a key parameter 
to determine the thickness of stellar disks.

\subsection{SFHs and $\Sigma_{\rm gas}-\Sigma_{\rm SFR}$ relations}
\label{sec:starformation}

\begin{figure}
\begin{center}
\FigureFile(80mm,80mm){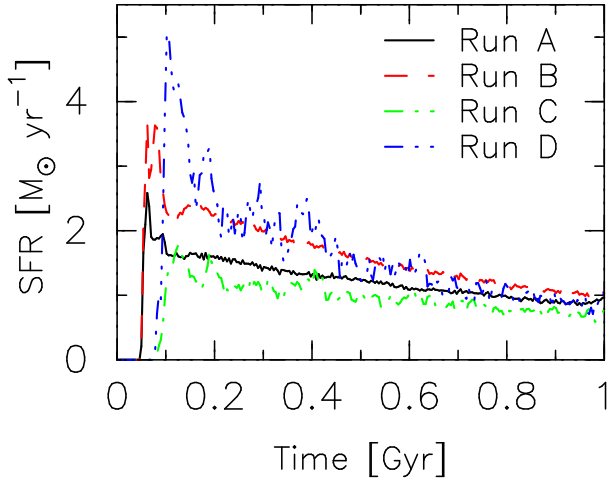}
\end{center}
\caption{Star formation histories in the simulations.
Solid, dashed, dot-dash, and dash-dot-dot-dot lines indicate Runs A, B, C, and D, respectively.
}\label{fig:SFH}
\end{figure}

Figure \ref{fig:SFH} compares the global SFHs in four runs (Runs A, B, C, and D).
In all runs, the SFHs are characterized by an initial 
rapid increase, followed by a gradual decrease.
The rapid increase continues for only $\sim 10^7~{\rm yr}$ because of SN feedback effects.
The evolutions after the first peak are described approximately by
an exponential decay with short-period spikes.
We find that the SFRs asymptotically decrease to $\sim 1-2~\Mo~{\rm yr^{-1}}$, 
which is close to the observational value of SFR in 
nearby spiral galaxies ($\sim 1~\Mo~{\rm yr^{-1}}$: \cite{James+2004}) and 
Galactic SFR ($\sim 3~\Mo~{\rm yr^{-1}}$: \cite{McKeeWilliams1997}).

It is worth pointing out that the difference in the global SFHs 
of Runs C and D is rather small,
despite a factor of $15$ ($= 0.5/0.033$) difference in $C_{*}$. 
\citet{TaskerBryan2006} performed test simulations by changing $C_{*}$ by a factor of ten
and found that it has little effect to the global SFHs. 
In other words, global SFHs are not directly proportional to $C_{*}$ 
when a high-density threshold is adopted.
They claimed that this is because the dynamical time in star forming regions is 
sufficiently shorter than that in galaxies.
As will be discussed in \S \ref{sec:fluidevolution}, 
we find that 
the contraction timescale of the gas is about 5 times longer than 
the local dynamical time and this timescale does not depend on the value of $C_*$.
Hence the global SFH is not directly proportional to $C_{*}$. 

\begin{figure}
\begin{center}
\FigureFile(80mm,80mm){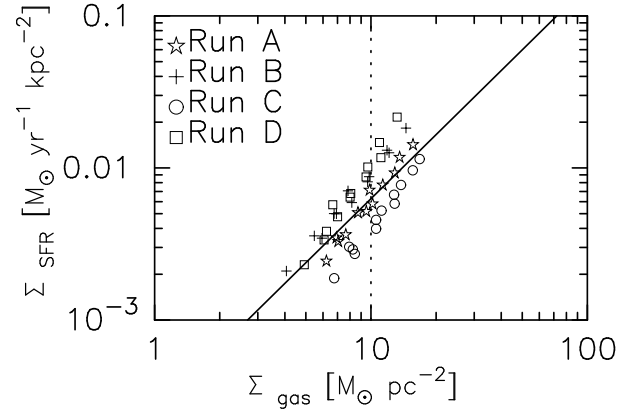}
\end{center}
\caption{Surface gas density and the surface star formation rate 
for Runs A, B, C, and D in three different epochs $t = 0.3,~0.5$, and  $1.0~{\rm Gyr}$.
The surface SFRs are computed using 
the surface densities of young star particles 
of which ages are shorter than the typical age of massive stars; $4.5 \times 10^{7}~{\rm yr}$.
The outer edges of surface densities correspond with 
the distance of the most distant young star particle from the galactic center
and the typical edges of the star forming regions are $R \simeq 10~{\rm kpc}$.
Four cylindrically averaged values with a constant radial interval are obtained from each run and each epoch. 
Stars, crosses, circles, and squares represent the sequences of Runs A, B, C, and D, respectively.
The solid line is a best fit from observations \citep{Kennicutt1998}.
}\label{fig:SigmaGas_SigmaSFR}
\end{figure}

Figure \ref{fig:SigmaGas_SigmaSFR} presents $\Sigma_{\rm gas}-\Sigma_{\rm SFR}$ relations 
for Runs A, B, C, and D.
All of our runs show similar relations, both in the slope and in the normalization, 
to the observational values (-1.4; solid lines in the figure),
although the range covered by our simulations is somewhat limited.
Slightly steeper slopes in Runs with a high critical density (i.e., Runs C and D)
than Run A is consistent with what \citet{WadaNorman2007} found in 
their theoretical model of global star formation in the ISM with log-normal density PDF.
We have shown that the different threshold models have very different ISM structures.
The distributions of newly formed star particles are also different, 
reflecting the structures of the ISM.
Nevertheless, as we have shown,
the observed $\Sigma_{\rm gas}-\Sigma_{\rm SFR}$ relation is
well reproduced in all our simulations.

\subsection{Phase structure of the ISM} \label{sec:phase}

In this section, we study the phase structures of the ISM in more detail.
The differences between the low-$n_{\rm th}$ and high-$n_{\rm th}$ models 
in phase diagram are studied in \S \ref{sec:rho-T}.
The distribution functions of gas mass as a function 
of density and temperature are also discussed.
Evolution of individual gas particles is analyzed in \S \ref{sec:fluidevolution}.
We find from the analysis that the evolution timescale of the ISM  
($1~{\rm cm^{-3}} < n_{\rm H} < 100~{\rm cm^{-3}}$) is typically $\sim 5~t_{\rm dyn}(n_{\rm H})$.
We show the PDFs of Runs C, D and E in \S \ref{sec:PDF}.
The high-density parts of PDFs in Runs C and D are well fitted by a log-normal function,
whereas that in Run E is a single power-law like form.
In our simulations, a form of PDF changes by the effects of star formation and SN feedback.

\subsubsection{Phase diagram} \label{sec:rho-T}
\begin{figure}
\begin{center}
\FigureFile(80mm,160mm){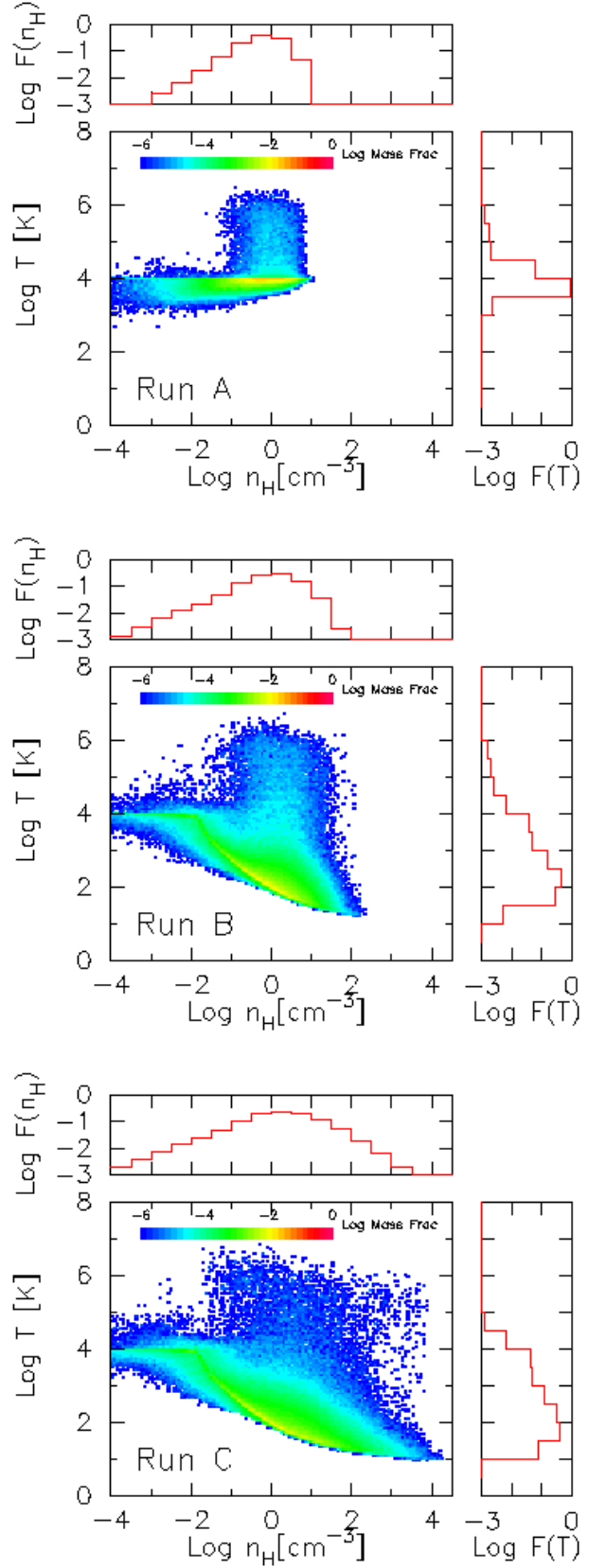}
\caption{
$\rho-T$ diagram for Runs A, B, and C (from the top panel to the bottom panel).
Whole plot regions are subdivide into $128 \times 128$ grids and  
each grid is colored by a mass fraction. 
Top and right histograms in each panel show mass fractions as functions of 
density and temperature.
}\label{fig:Phase}
\end{center}
\end{figure}

In figure \ref{fig:Phase}, we show global phase ($\rho-T$) diagrams 
of Runs A, B, and C at $t = 0.3~{\rm Gyr}$.
The phase diagram in Run D resembles that in Run C and thus we exclude it here.
Differences in phase distribution between Runs C and D are discussed in \S \ref{sec:PDF}.
Phase structures in different runs are very different.
Phase structures strongly depend on $n_{\rm th}$ and $T_{\rm cut}$.
As a consequence of the assumption ($T_{\rm cut} = 10^4~{\rm K}$),
Run A consists of a warm ($T \sim 10^4~{\rm K}$) and a hot ($T > 10^5~{\rm K}$) phase.
There is no cold phase because of the cut-off temperature of the cooling function, $T_{\rm cut}$,
whereas the hot phase is formed by the SN feedback.
In Run A, there is a density cut off at $\sim 10~{\rm cm^{-3}}$ because
star formation is rapidly consuming high-density gas  
and the gas pressure with  $T = 10^4~{\rm K}$ prevents further collapse.
The overall feature is similar to that in \citet{Stinson+2006}
and the assumption ($T_{\rm cut} = 10^4~{\rm K}$) 
is reasonable for coarse resolution ($m_{\rm SPH} \sim 10^{5-6}~\Mo$) runs (see figure \ref{fig:Jeans}).
Runs B and C show a clear multiphase structure with a cold gas.
The high-density gas in Run B becomes denser than that in Run A due to the low effective pressure.
The high-density tail extends to $\sim 100~{\rm cm^{-3}}$ in this case.
The high-$n_{\rm th}$ model has more dense and cold gas,
since the gas consumption due to star formation is occurred at $\sim 100~{\rm cm^{-3}}$.
The multiphase structure in Run C is similar to those obtained in previous 
high-resolution grid-base simulations (e.g., \cite{WadaNorman1999,Wada2001,TaskerBryan2006}).

The mass fractions as a functions of density $F(n_{\rm H})$ and temperature $F(T)$
are shown in top and right histograms in each panel of figure \ref{fig:Phase}.
In Run A, the peak of $F(n_{\rm H})$ is $n_{\rm H} \sim 1~{\rm cm^{-3}}$ 
and that of $F(T)$ is $T \sim 10^4~{\rm K}$.
The temperature peak is shifted to $\sim 100~{\rm K}$ in Run B.
The dominant component in temperature in Run C is also the cold phase gas ($T < 1000~{\rm K}$) 
as is also found in other high-resolution simulations of the ISM, 
i.e., $\sim 80-90$\% of mass in the ISM is in the cold phase
(i.e., \cite{RosenBregman1995,Wada2001,TaskerBryan2006}).
The peak of $F(n_{\rm H})$ is $n_{\rm H} \sim 1~{\rm cm^{-3}}$ 
and that of $F(T)$ is $T \sim 100~{\rm K}$.
In conclusion, the multiphase ISM in the high-$n_{\rm th}$ models has 
a large mass of reservoir around $1~{\rm cm^{-3}}$
for the star forming region ($n_{\rm H} > 100~{\rm cm^{-3}}$).
Thus the evolution of the reservoir should play a key role 
for the star formation in the multiphase ISM.
We further study the evolution of the gas from 
the reservoir to star forming regions in the next subsection.

\subsubsection{Detailed evolution of fluid elements in the multiphase ISM} \label{sec:fluidevolution}

\begin{figure}
\begin{center}
\FigureFile(80mm,160mm){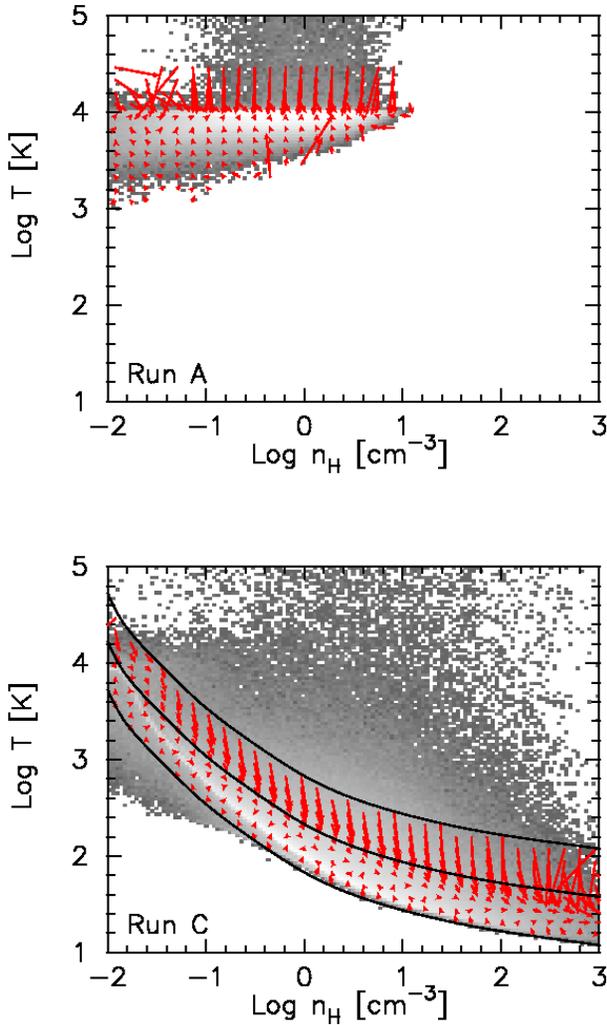}
\caption{
Evolution of SPH particles in each grid on phase diagram within $\Delta t = 1~{\rm Myr}$ 
at $t = 0.3~{\rm Gyr}$ in Runs A (top panel) and C (bottom panel).
Arrows indicate evolution of sampled grids on the phase diagram 
in $T < 10^5~{\rm K}$ for Run A and 
in $T_{\rm eq} < T < 10 \times T_{\rm eq}$ for Run C.
The heads of arrows indicate the median values of density and temperature 
after $\Delta t$ of evolution.
Black curves indicate the equilibrium temperature of 
the adopted cooling and heating with the solar abundance, 
$T_{\rm eq}$, and its families that $10^{0.5}~T_{\rm eq}$ and $10~T_{\rm eq}$, respectively.
Background gray scales indicate mass weighted distributions of gas on phase diagram.
} \label{fig:PhaseArrows}
\end{center}
\end{figure}

\begin{figure}
\begin{center}
\FigureFile(80mm,80mm){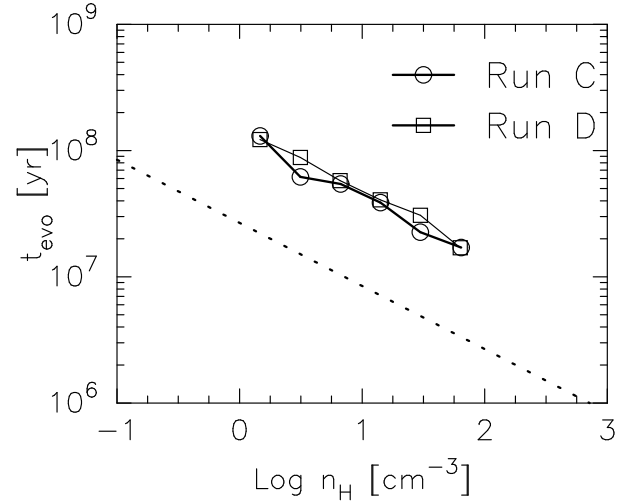}
\caption{
The e-folding times of density evolution 
for Run C (the thick solid line with circles) and Run D (the thin solid line with boxes).
See the text for the definition of the evolution timescale.
Dotted line represents the local dynamical time as a function of the density. 
} \label{fig:RhoFlux}
\end{center}
\end{figure}

Figure \ref{fig:PhaseArrows} shows the evolution of gas in the phase ($\rho-T$) diagram 
within $\Delta t \equiv 1~{\rm Myr}$ at $t = 0.3~{\rm Gyr}$ in Runs A and C.
There is a main stream toward higher densities between 
$T_{\rm eq}$ and $10^{0.5} \times T_{\rm eq}$ in Run C,
while we can not find any flow in that direction in Run A. 
We compute the mass flux on the phase diagram around the main stream,
between $T_{\rm eq}$ and $10 \times T_{\rm eq}$, toward high density 
with binning from $\log (n_{\rm H}) = 0$ to $\log (n_{\rm H}) = 2$ every $1/3~{\rm dex}$ in density.
The median values are adopted for the indicator of the density evolution in each bin.
Figure \ref{fig:RhoFlux} shows the e-folding (evolution) times in Runs C and D.
We define the evolution time that the density changes $e$ times 
by the evolution time of medians within $\Delta t$ interval on the phase diagram:
\begin{equation}
t_{\rm evo} \equiv \rho_{\rm m} / (\Delta \rho_{\rm m} / \Delta t)
\end{equation}
where $\rho_{\rm m}$ is initial values of median densities 
and $\Delta \rho_{\rm m}$ is density changes of $\rho_{\rm m}$ within $\Delta t$, respectively.
We then find that the evolution time is $t_{\rm evo} \sim 5~t_{\rm dyn}(n_{\rm H})$:
$t_{\rm evo}(n_{\rm H} \sim 100~{\rm cm^{-3}}) \sim 20~{\rm Myr}$ and
$t_{\rm evo} (n_{\rm H} \sim 1~{\rm cm^{-3}}) \sim 130~{\rm Myr}$.
The analysis on Run D shows almost the same results 
(compare the thin (Run D) and solid (Run C) lines in figure \ref{fig:RhoFlux}). 
Interestingly, the density dependence is proportional to
the local dynamical time: $t_{\rm evo} \propto \rho^{-1/2}$.
Comparison between Runs C and D indicates 
that the evolution timescale in the multiphase ISM is
independent of the star formation efficiency, $C_*$.

\begin{figure}
\begin{center}
\FigureFile(80mm,80mm){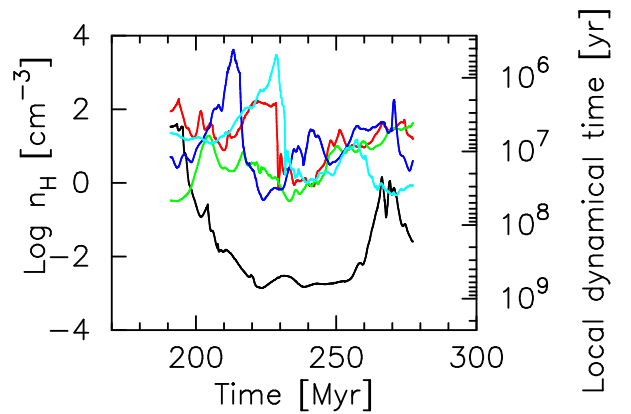}
\caption{
Evolution tracks of densities for 
randomly selected SPH particles from $190~{\rm Myr}$ to $280~{\rm Myr}$ in Run C.
Five colors denote five different particles.
Corresponding local dynamical time for density is shown in  the right side.
}\label{fig:DensityEvolution}
\end{center}
\end{figure}

Figure \ref{fig:DensityEvolution} shows density evolutions 
of several selected SPH particles in Run C. 
There are five loci of randomly selected SPH particles 
from $190~{\rm Myr}$ to $280~{\rm Myr}$.
All of the particles have the density changes of $\sim~5$ orders of magnitudes.
The evolutions are not monotonic but complex with many compressions and expansions.
The considerable drive mechanisms of the compressions are 
gravitational collapses and hydrodynamical shocks,
whereas those of the expansions are SNe and shear extensions. 
The global evolution of the ISM is expressed by the superpositions of above mechanisms.
Hence the mass flow from $\sim 1~{\rm cm^{-3}}$ to $\sim 100~{\rm cm^{-3}}$ 
is decelerated compared with that expected by only the gravitational collapse.

\subsubsection{The density PDF} \label{sec:PDF}

Shapes of PDF of the ISM give us an idea to 
connect the global structure of the ISM and star forming regions.
There have been a number of studies on this issue
(e.g., \cite{Vazquez-Semadeni1994,Scalo+1998,Vazquez-Semadeni+2000,Wada2001,
deAvillezMacLow2002,Kravtsov2003,deAvillezBreitschwerft2004,Slyz+2005,
WadaNorman2007,TaskerBryan2007,RobertsonKravtsov2007}).
When the fluid evolution is dominated by a random density change without any scale dependence,
it is expected that the volume weighted PDF becomes `log-normal' distribution 
\citep{Vazquez-Semadeni1994}.
In the multiphase ISM, there are many physical processes such as radiative cooling, FUV heating, 
star formation, SN feedback, hydrodynamical shock, and self-gravity.
It is unclear that whether the combination of these processes is either scale-invariant or not.
The expected form of a resultant PDF is also unclear.

\begin{figure}
\begin{center}
\FigureFile(80mm,160mm){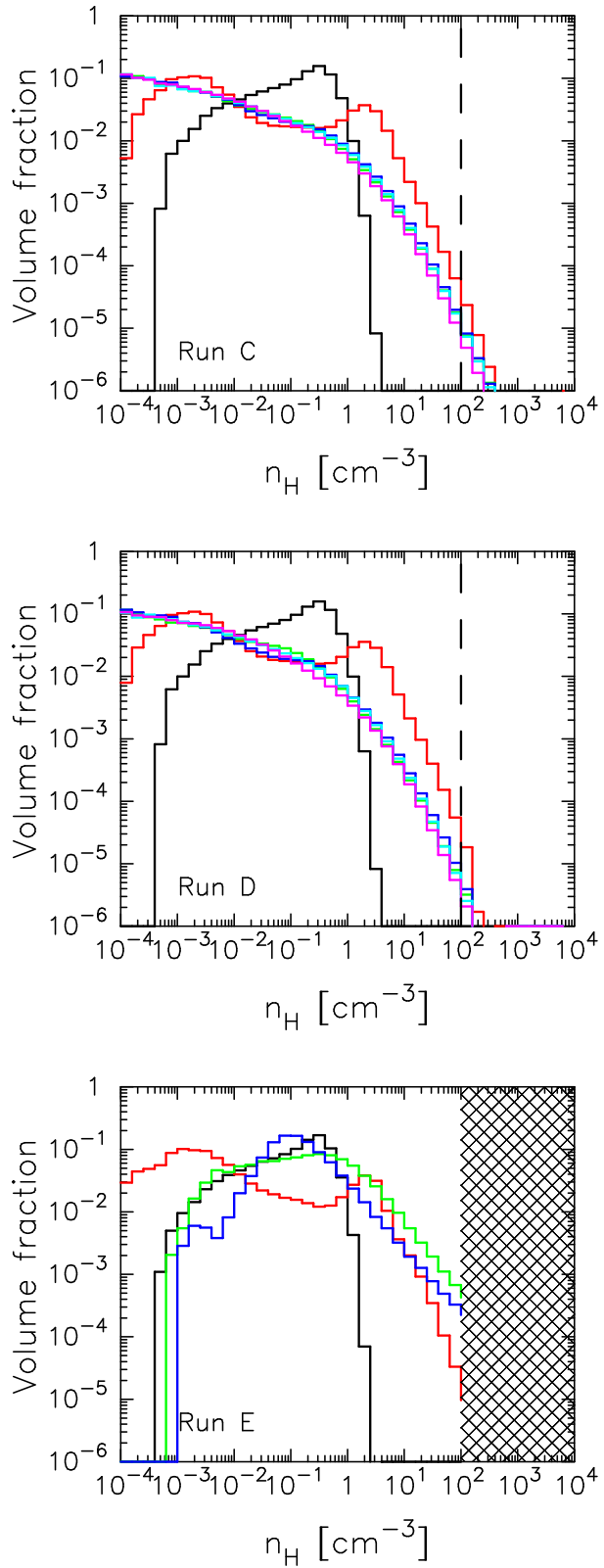}
\caption{Volume weighted probability distribution functions (PDFs) of 
Runs C, D, and E (from top to bottom) within $R < 10~{\rm kpc}$.
We calculate a volume fraction of an SPH particle by $V_i = m_i/\rho_i$.
The density interval is $0.2~{\rm dex}$. 
The PDFs are normalized to unity.
Runs C and D show six different epochs ($t = 0,~0.1,~0.2,~0.3,~0.5,$ and $1.0~{\rm Gyr}$),
while Run E shows first four epochs.
Black, red, green, blue, sky blue, and magenta 
indicate PDFs of different epochs $t = 0,~0.1,~0.2,~0.3,~0.5,$ and $1.0~{\rm Gyr}$, respectively.
The vertical dotted lines are corresponding 
with the star formation threshold density, $n_{\rm th} = 100~{\rm cm^{-3}}$.
The high-density region ($n_{\rm H} > 100~{\rm cm^{-3}}$) in Run E is hatched,
since the region has artificial mass stagnation due to insufficient mass resolution.
}\label{fig:PDF}
\end{center}
\end{figure}

Figure \ref{fig:PDF} shows the evolutions of PDFs in our SPH simulations.
Here, we plot only PDFs of Runs C, D, and E.
The evolutions of PDFs in these three runs are almost the same at the first phase,
although the quasi-static final shapes depend on models.
At initial state, the PDFs have uniform density distributions with 
a peak around $n_{\rm H} \sim 0.3~{\rm cm^{-3}}$.
The PDFs are smoothed out within first $\sim 0.2~{\rm Gyr}$.
Our PDFs in the three runs then become quasi-static 
as reported by 
\citet{WadaNorman2001}, \citet{Kravtsov2003}, and \citet{WadaNorman2007}.
High density tails ($n_{\rm H} > 1~{\rm cm^{-3}}$) in Runs C and D 
are in steady state and have log-normal forms (e.g., 
\cite{Vazquez-Semadeni1994,Vazquez-Semadeni+2000,Wada2001,Kravtsov2003,
WadaNorman2007,TaskerBryan2007}).
The high-density tail in Run E,
the model does not include either star formation or SN feedback, 
is asymptotic to a power-law form (green and blue lines).
The difference in high-density region comes from the fact that
the gas in the high-density region converts into stars and 
SN feedback blows out the surrounding dense gas in Runs C and D.
\citet{Slyz+2005} showed that a high-density tail of a PDF has 
a power-law like form in a self-gravity-dominated system,
although several authors argued that the log-normal PDF is the robust structures of multiphase ISM, 
regardless of the input physics \citep{WadaNorman2001,Kravtsov2003}.
Further investigation into details is required 
for the response of the input physics (e.g., UV radiation, \authorcite{SusaWada} in preparation) in the form of PDF.

\subsection{Convergence tests}\label{sec:resolution}

\begin{figure*}
\begin{center}
\FigureFile(160mm,160mm){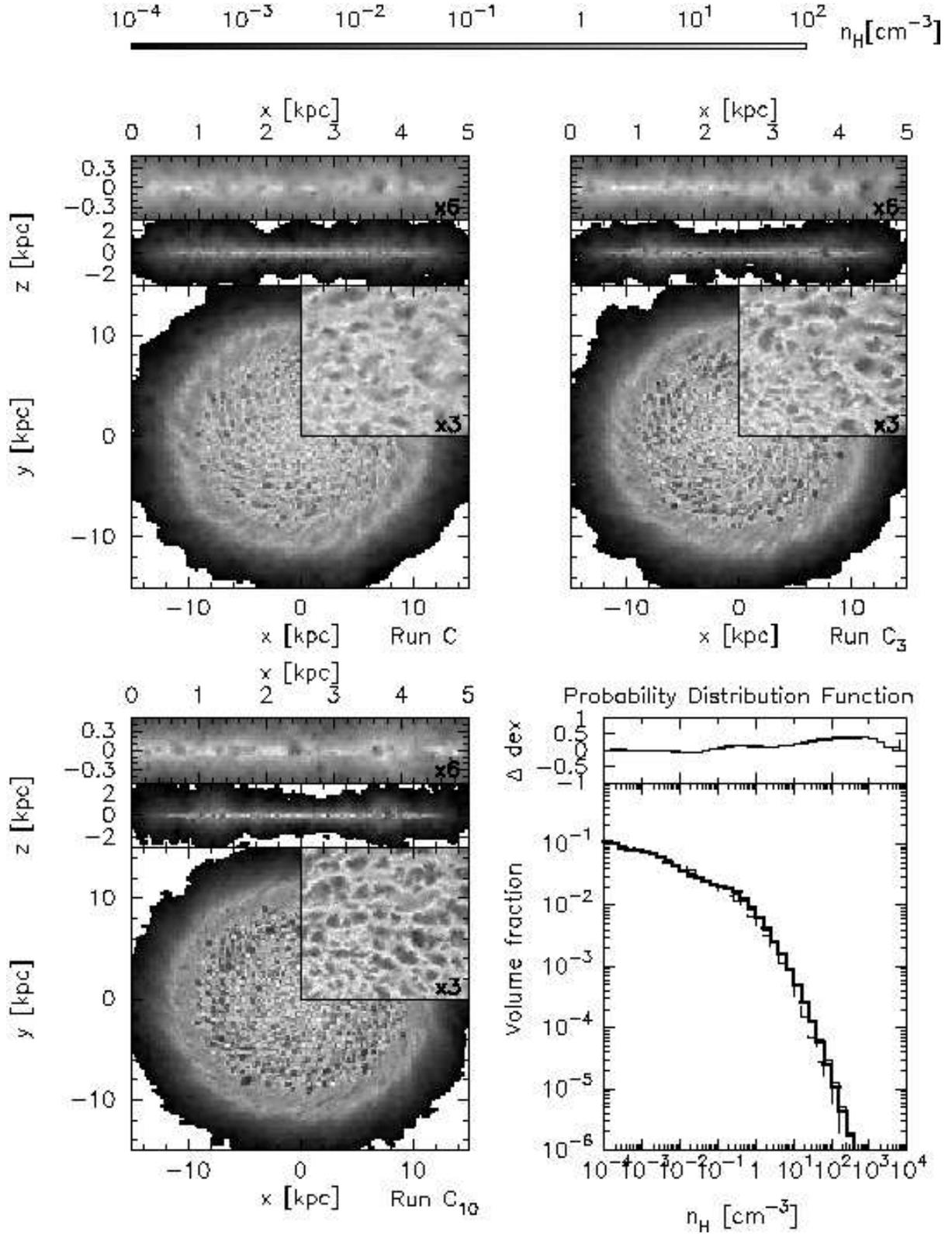}
\caption{
Sets of panels at tops and left-bottom are gas density maps for three different mass resolutions 
($N = 10^6, 3 \times 10^6$, and $10^7$, respectively).
The right-bottom panel shows PDFs of three runs at $t = 0.3~{\rm Gyr}$.
Thin dashed, thick solid, and thin solid lines indicate simulations using 
$N = 10^6, 3 \times 10^6$, and $10^7$, respectively.
PDF residuals of Runs C$_3$ and C$_{10}$ from C is shown in the upper portion of the panel.
}
\label{fig:GasResolution}
\end{center}
\end{figure*}

\begin{figure}
\begin{center}
\FigureFile(80mm,80mm){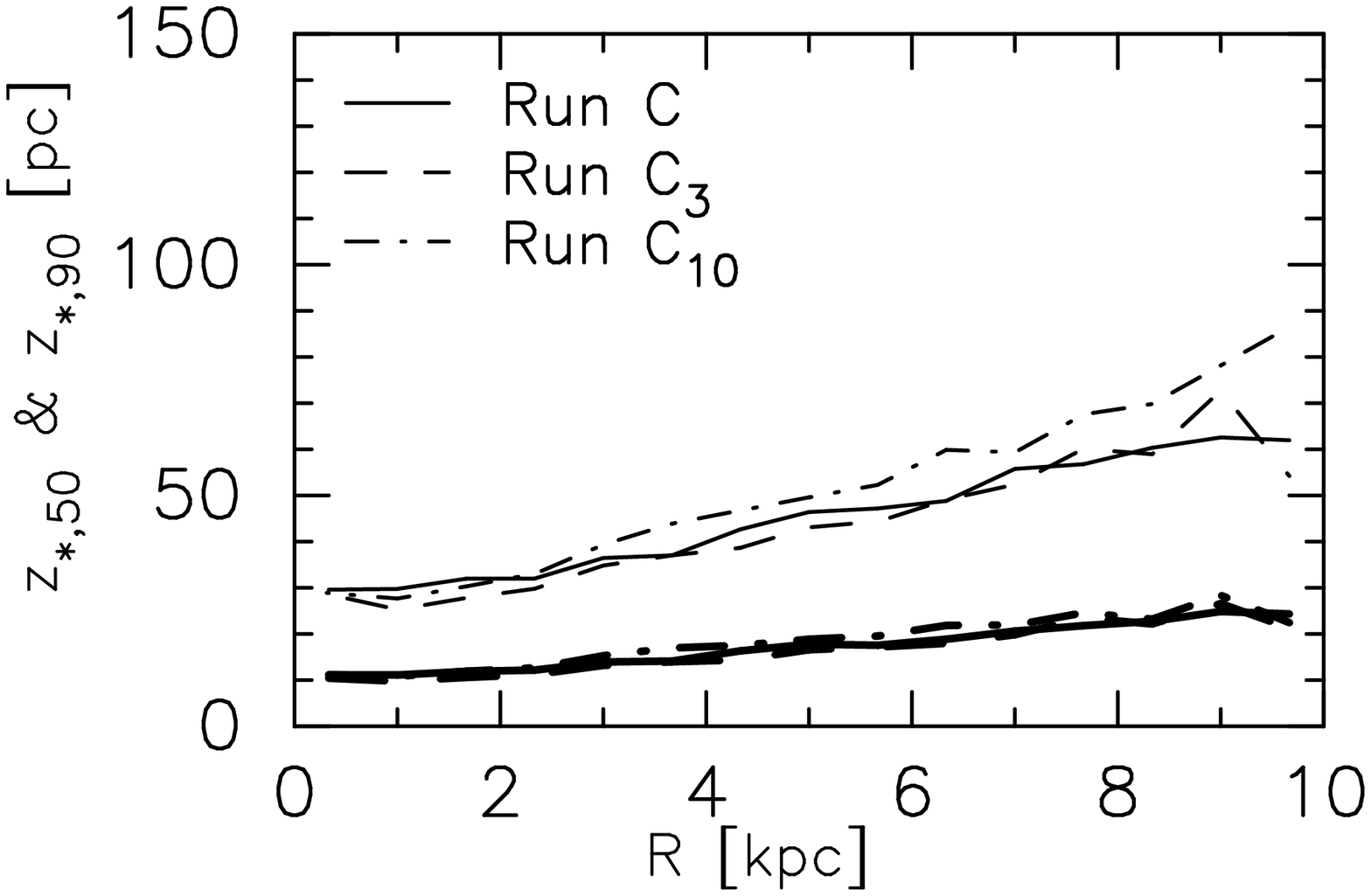}
\caption{Same as figure \ref{fig:zrstar}, but for Runs C, C$_{3}$, and C$_{10}$.
}\label{fig:ConvergenceStellar}
\end{center}
\end{figure}

Figure \ref{fig:GasResolution} shows density maps in the simulations using
$N = 10^6, 3 \times 10^6$, and $10^7$ (Runs C, C$_3$, and C$_{10}$) at $t = 0.3~{\rm Gyr}$.
At the time, the numbers of particles in all of runs increase $\sim 30~\%$ from the initial states
due to star formation.
The density distributions indicate that differences among 
the three runs with different mass resolutions are very little.
These gas disks have almost the same sizes of filaments and voids.
The complexness of the gas disks is almost the same degree.
The statistical structures of the multiphase ISM, PDFs, 
in the three runs are also similar to one another
(see the lower-right panel in figure \ref{fig:GasResolution}).
Differences are found in the high-density tails from residuals.
However the residuals are only factor of three at most.
Figure \ref{fig:ConvergenceStellar} shows vertical thicknesses of stellar disks.
These thickness are almost the same for all runs.
Thus we consider that our results are roughly converged in above mass resolutions,
for our selected values for $n_{\rm th}$ and $T_{\rm cut}$.

\section{Summary and Discussion} \label{sec:discussion}
\subsection{Importance of the density threshold for star formation}

In many studies, numerical models of star formation in galaxy formation have been 
calibrated to be consistent with the observational 
$\Sigma_{\rm gas}-\Sigma_{\rm SFR}$ relation.
For example, \citet{Stinson+2006} successfully reproduced the relation
with the threshold density of $n_{\rm th} = 0.1~{\rm cm^{-3}}$.
In order to reproduce the relation within simulated disk galaxies, 
\citet{SpringelHernquist2003} and \citet{SchayeDallaVecchia2007} 
explicitly involved the Schmidt-Kennicutt relations in star formation models in the ISM.
\citet{Kravtsov2003} showed that a high-density threshold ($n_{\rm th} = 50~{\rm cm^{-3}}$) 
with a constant star formation time ($= 4~{\rm Gyr}$) reproduce 
the $\Sigma_{\rm gas}-\Sigma_{\rm SFR}$ relation.
\authorcite{TaskerBryan2006} (\yearcite{TaskerBryan2006}, \yearcite{TaskerBryan2007})
reported the comparison of 
the low-$n_{\rm th}$ and high-$n_{\rm th}$ star formation model.
Their simulations revealed that both models are 
able to reproduce the observed $\Sigma_{\rm gas}-\Sigma_{\rm SFR}$ relation.
We examined two models,
a low-$n_{\rm th}$ model ($n_{\rm th} = 0.1~{\rm cm^{-3}}$: Runs A and B) 
and a high-$n_{\rm th}$ model ($n_{\rm th} = 100~{\rm cm^{-3}}$: Runs C and D).
Our results also show that both models can reproduce 
the $\Sigma_{\rm gas}-\Sigma_{\rm SFR}$ relation (see figure \ref{fig:SigmaGas_SigmaSFR}).

We argue that we should generate ``stars'' above
the physical density of real star forming regions such as GMCs or molecular cores
to investigate the detailed structure and evolution of disk galaxies.
In this paper, we highlighted on the ISM structure and the distribution of newly formed star particles.
We found that (1) only the high-$n_{\rm th}$ model reproduces
the complex, inhomogeneous, and multiphase ISM structures 
(see figures \ref{fig:RunsABCD} and \ref{fig:RunsABCDStars}),
where the cold gas dominates in mass (see figure \ref{fig:Phase}).
These natures are well comparable with other studies of the ISM:
the geometrically complex and inhomogeneous structures of the ISM 
(e.g., \cite{RosenBregman1995,WadaNorman1999,deAvillez2000,deAvillezBerry2001,
TaskerBryan2006,TaskerBryan2007}),
three phases structures of the ISM where the cold mass dominates 
(e.g., \cite{McKeeOstriker1977,Myers1978,RosenBregman1995,
Wada2001,TaskerBryan2006,TaskerBryan2007}).
The log-normal PDF in the ISM is also found in the high-$n_{\rm th}$ models
(see figure \ref{fig:PDF} and figure \ref{fig:GasResolution})
although the origin of the log-normal shape appear to be different from that in previous studies 
(e.g., \cite{Vazquez-Semadeni1994,Vazquez-Semadeni+2000,
Wada2001,Kravtsov2003,WadaNorman2007,TaskerBryan2007}).
(2) Only the high-$n_{\rm th}$ models can reproduce observationally reported
scale heights of gas disks and young star forming regions 
as shown in figures \ref{fig:zrgas} and \ref{fig:zrstar}.
Therefore we emphasize that the density threshold for star formation model is 
the key parameter to model realistic three-dimensional structures of galaxies, 
especially gas and stellar disk structures.
We have to choose the star forming gases as the physical one for this purpose.
It is necessary to solve energy equation for much lower temperature gas than $10^4~{\rm K}$
to resolve the high-density gas.
This requires higher mass resolution than those used 
in previous simulation of galaxy formation.

\subsection{Weak dependence on star formation efficiency}

Runs C and D differ in the values of $C_*$.
The results are, however, similar in terms of 
the ISM structure and stellar disks (figures \ref{fig:RunsABCD} and \ref{fig:RunsABCDStars}),
the SFHs (figure \ref{fig:SFH}), the $\Sigma_{\rm gas}-\Sigma_{\rm SFR}$ relation 
(figure \ref{fig:SigmaGas_SigmaSFR}), and the phase distribution of the ISM (figure \ref{fig:PDF}).
Why do the simulations show similar results? 
As shown in figure \ref{fig:Phase}, a large fraction of the gas 
exist at around $\sim 1~{\rm cm^{-3}}$ and 
it behaves as the reservoir of the star forming gas.
The mass supply timescale from the reservoir to the star forming region 
determines the global star formation rate
in the model that adopts the high-density threshold ($n_{\rm th} = 100~{\rm cm^{-3}}$).
From figure \ref{fig:RhoFlux},
we find that the timescale is $\sim 5~t_{\rm dyn}(n_{\rm H})$ and
the timescale is independent of the values of $C_*$.
Hence $C_{*}$ has only weak effects on the global features of the ISM 
and stellar disks formed from the ISM.

One of the most important advantages of the high-$n_{\rm th}$ model is that 
global SFR is not determined by the local quantities but by the global state of the ISM.
The exact value of $C_*$ is unimportant in galactic scale simulations,
hence we can effectively avoid an uncertainty in the star-formation model. 
The remaining weak dependences of $C_*$ 
would vanish when we adopt even higher threshold density for star formation
together with higher resolution.

When we tune the value of $C_*$ in low-$n_{\rm th}$ models, we can obtain
the similar global star formation properties in disk galaxies
regardless of whether we resolve the detailed structure of the ISM or not.
This finding provides a support for star formation models in lower
resolution cosmological simulations, which is similar to Run A
in this paper.
We will investigate whether it is also true for starburst galaxies in
forthcoming papers.

\ 

We thank the anonymous referee for his/her fruitful comments and suggestions.
The author (TRS) thanks Nozomu Kawakatu and Junichi Baba for helpful discussions.
Numerical computations were carried out on GRAPE system (project ID:g06a15/g07a19)
at the Center for Computational Astrophysics, CfCA, 
of the National Astronomical Observatory of Japan.
This project is supported by Grant-in-Aids for Scientific Research (17340059) of JSPS
and Molecular-Based New Computational Science Program of
National Institutes of Natural Sciences.

\end{document}